\documentclass[reprint,prb]{revtex4-1}
\usepackage{hyperref}
\usepackage{upgreek}
\usepackage[latin1]{inputenc}
\usepackage{subsupscripts}
\usepackage{subscript}

\usepackage{graphicx}
\usepackage{dcolumn}
\usepackage{bm}
\usepackage{color}

\begin{document}

\preprint{}

\title{Influence of interactions with non-condensed particles on the coherence of a 1D polariton condensate}
\author{Johannes Schmutzler}
\affiliation{Experimentelle Physik 2, Technische Universit\"at Dortmund, \mbox{D-44221 Dortmund, Germany}}
\author{Tomasz Kazimierczuk}
\affiliation{Experimentelle Physik 2, Technische Universit\"at Dortmund, \mbox{D-44221 Dortmund, Germany}}
\author{\"Omer Bayraktar}
\affiliation{Experimentelle Physik 2, Technische Universit\"at Dortmund, \mbox{D-44221 Dortmund, Germany}}
\author{Marc A{\ss}mann}
\affiliation{Experimentelle Physik 2, Technische Universit\"at Dortmund, \mbox{D-44221 Dortmund, Germany}}
\author{Sebastian Brodbeck}
\affiliation{ Technische Physik, Physikalisches Institut, Wilhelm Conrad R\"ontgen Research Center for Complex Material Systems,
Universit\"at W\"urzburg, D-97074 W\"urzburg, Germany}
\author{Martin Kamp}
\affiliation{ Technische Physik, Physikalisches Institut, Wilhelm Conrad R\"ontgen Research Center for Complex Material Systems,
Universit\"at W\"urzburg, D-97074 W\"urzburg, Germany}
\author{Christian Schneider}
\affiliation{ Technische Physik, Physikalisches Institut, Wilhelm Conrad R\"ontgen Research Center for Complex Material Systems,
Universit\"at W\"urzburg, D-97074 W\"urzburg, Germany}
\author{Sven Höfling}
\email[Current address: ]{SUPA, School of Physics and Astronomy, University of St Andrews, St Andrews, KY16 9SS, United Kingdom.}
\affiliation{ Technische Physik, Physikalisches Institut, Wilhelm Conrad R\"ontgen Research Center for Complex Material Systems,
Universit\"at W\"urzburg, D-97074 W\"urzburg, Germany}
\author{Manfred Bayer}
\affiliation{Experimentelle Physik 2, Technische Universit\"at Dortmund, \mbox{D-44221 Dortmund, Germany}}

\date{28 February 2014}

\begin{abstract}
One-dimensional polariton condensates (PoCos) in a photonic wire are
generated through non-resonant laser excitation, by which also a
reservoir of background carriers is created. Interaction with this
reservoir may affect the coherence of the PoCo, which is studied here
by injecting a condensate locally and monitoring the coherence along
the wire. While the incoherent reservoir is mostly present within
the excitation laser spot, the condensate can propagate ballistically
through the wire. Photon correlation measurements show that far from
the laser spot the second order correlation function approaches unity
value, as expected for the coherent condensed state. When
approaching the spot, however, the correlation function increases up
to values of 1.2 showing the addition of noise to the emission due
to interaction with the reservoir. This finding is substantiated by
measuring the first order coherence by a double slit
experiment, which shows a reduced visibility of interference at the
excitation laser spot.
\end{abstract}



\maketitle
\section{introduction}
After the demonstration of Bose-Einstein-Condensation (BEC) of
alkali atoms \cite{Davis1995,Anderson1995} the efforts to observe
condensation phenomena in condensed matter were intensified, because
the small mass of candidate excitations promises high condensation
temperatures. Particularly promising are polaritons in semiconductor
microcavities in the strong coupling regime\cite{Weisbuch1992}: due
to their photon admixture the polariton mass is extraordinarily small,
enabling condensation under ambient conditions. The efforts soon rendered
success  with demonstration of polariton condensation in several semiconducting materials. \cite{Deng2002,Kasprzak2006,Balili2007,Christopoulos2007,Franke2012} In the
meantime unambiguous criteria to distinguish polariton lasing by a
polariton condensate (PoCo) from photon lasing have been worked out.
\cite{Bajoni2007,Sanvitto2010,Assmann2011,Tempel2012a,Schneider2013}

A polariton laser promises low power consumption as it operates
without the need for population inversion.\cite{Imamoglu1996} For
practical applications the carriers have to be injected
non-resonantly with significant excess energy, e.g., by electrical currents, as demonstrated very
recently.\cite{Schneider2013,Bhattacharya2013} Consequently, during relaxation a broad distribution of background carriers is generated. This might limit the performance of such
a device in terms of coherence of the emission due to
interaction between condensed polaritons and uncondensed
particles. Comparative linewidth measurements on 2D PoCos \cite{Askitopoulos2013} indicated  already, that the separation of reservoirs carriers from the PoCo may improve the coherence properties.

PoCos in 2D cavity structures\cite{Deng2007,Lai2007,Belykh2013} are affected not only by interaction with background carriers but undergo also considerable scattering among the polaritons. Such scattering is elastic in that the energy maintains in the polariton system. In 2D cavities the phase space of possible scattering events is rather large so that the bare effect of background carriers on the PoCo coherence is hard to assess. This is in particular the case when the coherence is studied underneath the laser spot only as in Refs. \onlinecite{Deng2007,Lai2007,Belykh2013}. Promising in this respect are photonic wire structures in which the one-dimensionality suppresses polariton scattering due to the reduced phase space accessible for quasi-elastic scattering. Propagation several $10~\upmu\mbox{m}$ away from the excitation laser spot has been demonstrated in these structures.\cite{Wertz2010}

In this report we present a spatially resolved study of the
coherence properties of a laser-excited 1D PoCo. To that end, we use
two complementary experimental techniques, namely a second order
correlation measurement\cite{Wiersig2009, Assmann2010} as well as a
Young's double-slit experiment. Both approaches give
evidence for a reduced coherence of the PoCo when background carriers
are present. On the other hand, polaritons propagating out of
the excitation spot maintain their coherence, so that with
increasing separation coherence is established. We attribute this loss of coherence to interaction between background carriers and the PoCo. To our opinion two different effects contribute to the interaction: (i)  The Coulomb-potential mediated by the background carriers and (ii) nonresonant scattering between background carriers and polaritons. However, a detailed evaluation of the exact contributions of these effects is beyond the scope of this report. 

This manuscript is structured as follows: In Sec. II the investigated sample as well as the used experimental techniques are described. Here, special attention is attributed to the second order correlation measurements using a streak camera and the corresponding data analysis. This is followed by a presentation of our experimental results in Sec. III. Finally, a conclusion and an outlook for further experiments is given in Sec. IV.

\section{Experimental details}

We investigate a GaAs-based $\lambda/2$-microcavity with an experimentally determined Q factor of about $10000$. The design of the sample is as follows:
Three stacks of four GaAs quantum wells are placed in the three central antinodes of the electric field confined by two distributed Bragg reflector (DBR) structures in a $\uplambda$/2-cavity. The quantum wells are $13~\mbox{nm}$ in thickness and separated by $4~\mbox{nm}$ thick barrier layers of AlAs. The upper (lower) DBR structure consists of 23 (27) alternating layers of Al\textsubscript{0.2}Ga\textsubscript{0.8}As and AlAs. The interaction of the cavity field with the exciton resonance of the 12 contained GaAs quantum wells leads to a Rabi
splitting of about $10.5~\mbox{meV}$.
Photonic wires are fabricated by lithography and etching. A wire with the following parameters is used: The exciton-cavity detuning $\delta=E_C-E_X=-15.1~\mbox{meV}$.\footnote[1]{For convenience the exciton-photon detuning is defined as difference in energy between
the lowest confined, $i=0$ cavity subbranch and the uncoupled
quantum well exciton.} The wire length $L=100~\upmu\mbox{m}$ and
the wire width $W=5~\upmu\mbox{m}$.

The sample is mounted in a helium-flow cryostat, measurements are
performed at $10~\mbox{K}$. For non-resonant optical excitation a
femtosecond-pulsed Titanium-Sapphire laser (repetition rate
$75.39~\mbox{MHz}$) with central wavelength at $740~\mbox{nm}$
($1675~\mbox{meV}$) is used. The laser beam is focused under
normal incidence onto the sample, the shape of the spot is Gaussian
and about $2~\upmu\mbox{m}$ in diameter.

The emission from the sample is collected using a microscope objective (numerical aperture $0.42$); the far field emission is studied by imaging the Fourier plane of the objective onto the entrance slit of a monochromator. For detection a liquid nitrogen-cooled CCD-camera is used. For real-space imaging the photonic wire is magnified by a factor of $87.5$ and projected onto the entrance slit of a monochromator. 

To study the spatial coherence of the PoCo along the wire we have performed Young's double-slit experiment. Therefore, we magnify the emission
from the sample by a factor 100 onto four different double slits.
Thereby the spatial coherence between two small areas of diameter
$b=0.4~\upmu\mbox{m}$ with distances of $a=1.25~\upmu\mbox{m,
}2.5~\upmu\mbox{m, }5~\upmu\mbox{m and }7.5~\upmu\mbox{m}$ can be
investigated.  The interference fringes are recorded with the CCD camera behind the monochromator. Contrary to previous reports
\cite{Deng2007,Lai2007,Wertz2010,Belykh2013}, we have
chosen the location of the slit center $d$ with respect to the
excitation laser spot as an additional variable to investigate the
spatial coherence. $d=0$ corresponds to the situation of the double
slit placed symmetrically with respect to the excitation laser spot.
From observed interference fringes the visibility
$V=\frac{I_{max}-I_{min}}{I_{max}+I_{min}}$ within a spectral range of $0.5~\mbox{meV}$ is calculated, which is
used as measure for spatial coherence.

\begin{figure}[!t]
\centering
\includegraphics[width=\linewidth]{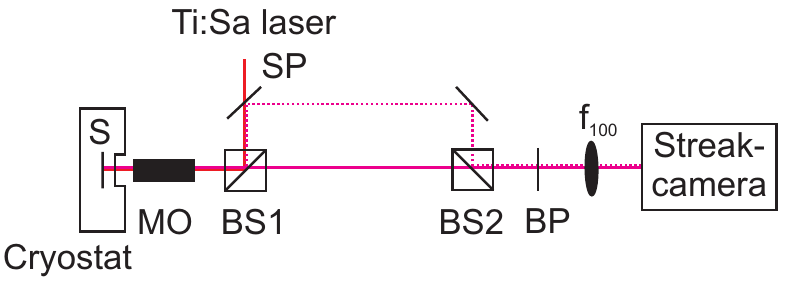}
\caption{(Color online) Schematic setup for the second order correlation measurement: BP, bandpass filter (FWHM 1~nm); BS1, BS2, beam splitters; f\textsubscript{100}, lens with 100 mm focal length; MO, microscope objective (focal length 4~mm); S, sample; SP, shortpass filter (cutting wavelength 800~nm). Note: Emission path indicated by the dashed line is delayed by 72.5~ps due to optical path length difference.} \label{Fig1}
\end{figure}

For the measurement of $g_2(\tau)$ we have slightly modified the streak camera setup described in Ref. \onlinecite{Assmann2010}. A drawback of the experimental approach presented there lies in photon reconstruction errors especially for short time delays $\tau<1-2~\mbox{ps}$ of the built-in streak camera routine in the single photon counting mode. This prevents a direct determination of the $g_2$-function for $\tau=0$ which can only be extrapolated from values of the $g_2$-function for larger $\tau$. Due to these photon reconstruction errors one can speak of a dark time in the order of $2~\mbox{ps}$. A similar problem occurs when measuring photon-statistics using avalanche photodiodes which exhibit a dark time in the order of $100~\mbox{ns}$. To circumvent this problem two avalanche photodiodes in conventional Hanbury-Brown-Twiss (HBT) setups are used. Similar as in a HBT setup, we can use our streak-camera actually as two detectors by the following approach:

We split the emission of the photonic wire  into two different optical paths delayed by $72.5~\mbox{ps}$ in time, thereby giving access to $g_2(\tau'=72.5~\mbox{ps})=g_2(\tau=0)$, where $\tau$ reflects the real time delay between the detection of two photons and $\tau'$ the time delay due to the artificial time delay given by different optical path lengths. 
Therefore two 50:50 beamsplitters and a shortpass (SP) filter with cutting wavelength at $800~\mbox{nm}$ are used (Fig. \ref{Fig1}). The SP-filter provides transmission of the  excitation laser as well as reflection of the investigated polariton-emission from the sample. Both emission patterns are magnified by a factor of 25 onto the entrance slit of a streak camera equipped with an additional horizontal deflection unit. The temporal resolution of the setup is approximately $2~\mbox{ps}$. Spectral sensitivity is provided by a bandpass filter with a Full Width at Half Maximum (FWHM) of $1~\mbox{nm}$. 

\begin{figure*}[!htb]
\centering
\includegraphics[width=\linewidth]{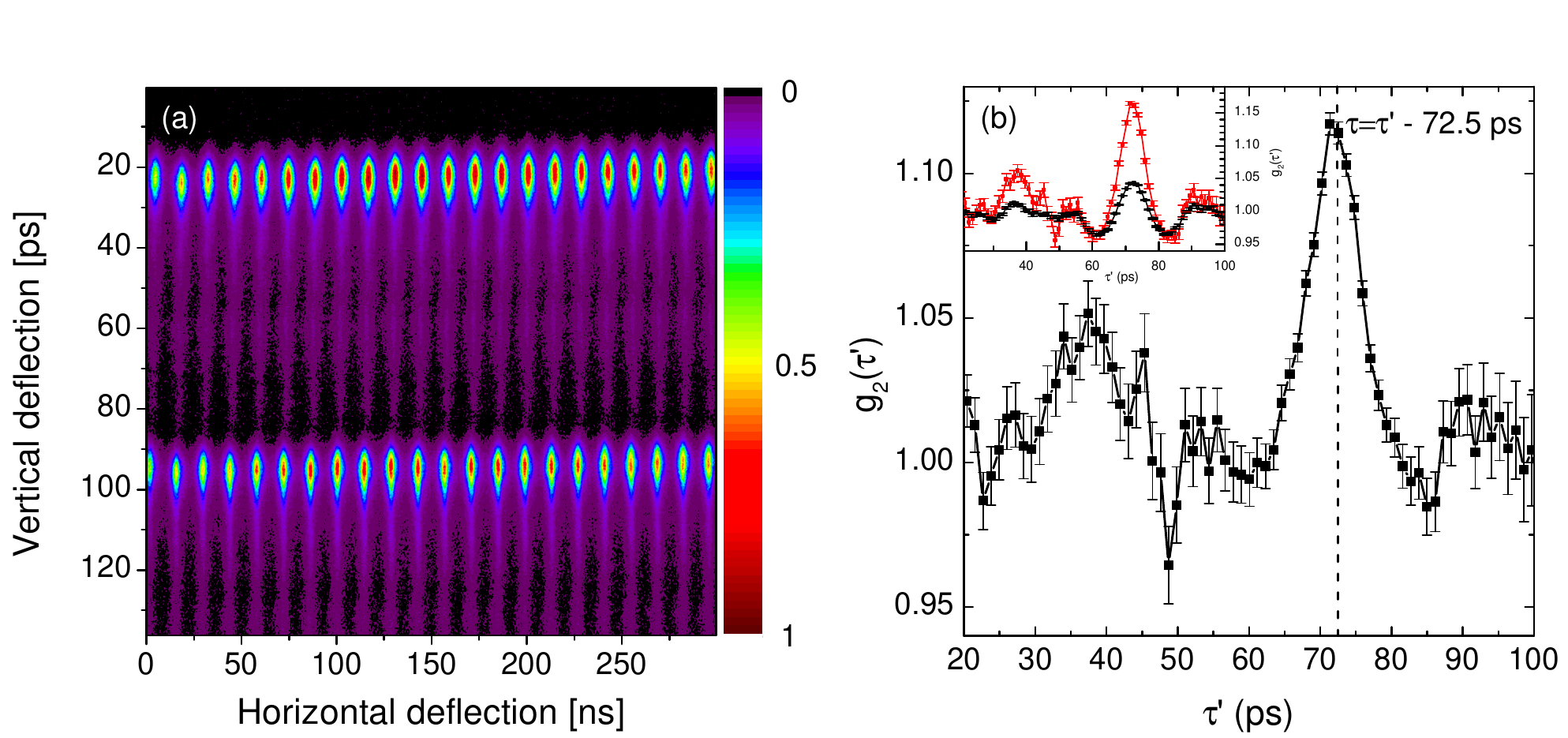}
\caption{(Color online) (a) Typical  image integrated over 100000 frames when probing a $3~\upmu\mbox{m}$ section of the photonic wire using a horizontal deflection time of $300~\mbox{ns}$. (b) Normalized $g_2(\tau ')$-function. Due to the artificial time delay between both profiles, $\tau '=72.5~\mbox{ps}$  corresponds to $\tau =0~\mbox{ps}$. Inset: Red line, calculated $g_2(\tau ')$-function when correlating photons within the same streak; black line, average of the correlation functions when correlating photons between different streaks.}  
\label{Fig2}
\end{figure*}

For the second order correlation measurement we select emission regions of $3~\upmu\mbox{m}$ width along the wire, using a vertical slit, and use a horizontal deflection time of  $300~\mbox{ns}$ per screen. For a reliable signal to noise ratio 100~000 frames are recorded, Fig. \ref{Fig2}~(a) shows a typical integrated image over 100~000 frames. Every frame consists of 22 streaks and each streak corresponds to one single excitation pulse. Every photon within one frame is sorted into different streaks and second order correlation functions are calculated as described in Ref. \onlinecite{Assmann2010}. A typical example for such a $g_2$-function is shown by the red line in the inset of  Fig. \ref{Fig2} (b). However, especially for short pulses the $g_2$-function can be distorted by jitter-effects as described in Ref. \onlinecite{Assmann2010}, which are indicated by $g_2$-values significantly below $1$. To account for these effects we average the $g_2$-functions between several combinations of different streaks which is indicated by the black line in the inset of Fig. \ref{Fig2}~(b).  Since neighboring streaks are separated by $13.2~\mbox{ns}$ in time, the shape of this curve does not reflect second order correlation of the emission from the sample, but jitter arising from our streak camera system. By dividing the $g_2$-function of photons within the same streak by the average of the $g_2$-functions of photons between different streaks we can separate jitter from second order correlation of the emission from the sample. Fig.~\ref{Fig2}~(b) gives a typical example of such a normalized $g_2$-function. Here $g_2(\tau'=72.5~\mbox{ps})$ corresponds to $g_2(\tau=0)$ due to the time delay between the two emission profiles. The additional peak for $\tau'=40~\mbox{ps}$ is probably caused by the tail of the pulse as it can be seen in Fig. \ref{Fig2}~(a). 

\section{Results and discussion}

\begin{figure*}[!htb]
\centering
\includegraphics[width=\linewidth]{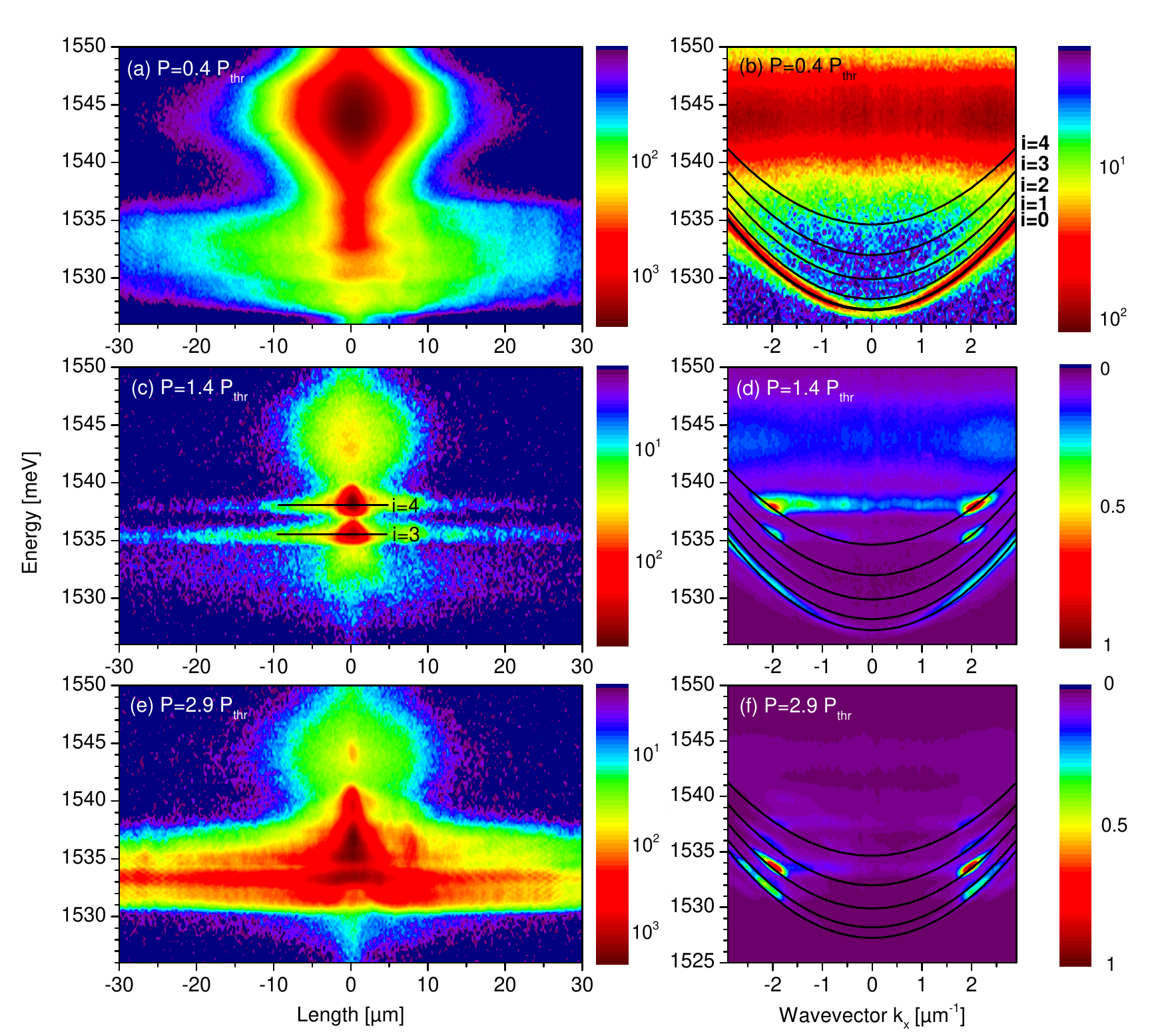}
\caption{(Color online) Real-space images (a), (c) and (e) and
Fourier-space images (b), (d) and (f) of the photonic wire at
different excitation power levels. (a): Also below threshold
propagation effects of the LPs, located in the energy range of
$1527-1537~\mbox{meV}$, are observed. The strong emission
centered at $1544~\mbox{meV}$ arises from the QWE. (b): Dispersion
of several LP subbranches can be distinguished. Black lines
correspond to calculated curves. (c) and (d): Slightly above
threshold the main emission is from the $i=3$-, and
$i=4$-subbranches. (e) and (f): For further increased excitation
power levels the main emission is shifted toward lower LP
subbranches. The most intense emission occurs at
$|k_x|\approx2~\upmu\mbox{m}^{-1}$.} \label{Fig3}
\end{figure*}

\subsection{Real-space and Fourier-space spectroscopy}
For identification of the different subbranches of the lower
polariton \cite{Kuther1998} in the photonic wire and evaluation of
the PoCo propagation we perform real-space and Fourier-space
spectroscopy at different excitation powers.

Figs. \ref{Fig3} (a) and (b) show the corresponding
images for an excitation power below threshold. Here,
the excitation laser spot is located in the center of the photonic
wire. Several dispersion curves of LP subbranches can be
distinguished in panel (b) and allocated to different photonic wire
subbranches. The most intensive mode corresponds to the
i=0-subbranch, also weak signatures of the i=2-, i=3- and
i=4-subbranches are seen [Fig.~\ref{Fig3}~(b)]. The reason for the
weak signals from higher subbranches is the orientation of the
photonic wire parallel to the entrance slit of the monochromator,
leading to detection within a small $k_y$-space range
$|k_y|<0.12~\upmu\mbox{m}^{-1}$. For this detection geometry
the emission in Fourier-space is dominantly contributed by the
ground mode, due to the symmetric, nodeless mode pattern perpendicular to the wire axis.\cite{Kuther1998,Lecomte2013} The strong emission centered at
$1544~\mbox{meV}$ is attributed to the bare uncoupled quantum well
exciton (QWE). This QWE photoluminescence is emitted mostly through
the edge of the wire. In real-space the confined LP modes show up as
several emission peaks below the QWE in the energy range of
$1527-1537~\mbox{meV}$ [Fig.~\ref{Fig3}~(a)].
Already below threshold, propagation of the LPs along the wire is
observed [Fig.~\ref{Fig3}~(a)], which is extended compared to the
exciton due to the light polariton mass.

At threshold significant changes of the emission patterns occur both
in real- [Fig. \ref{Fig3} (c)] and in Fourier-space [Fig. \ref{Fig3}
(d)] due to PoCo formation. The PoCo emission is most pronounced from
the i=3- and i=4-subbranches with the main emission at wavevectors
of $|k_x|\approx2~\upmu\mbox{m}^{-1}$. This can be attributed to
conversion of potential energy mediated by background carriers
within the excitation laser spot into kinetic energy, and to pair
scattering effects.\cite{Wertz2010} When the
excitation power is further increased, the main emission shifts to
lower subbranches [Fig. \ref{Fig3} (e) and (f)] and the propagation
along the photonic wire becomes much more pronounced. The emission in
real-space broadens, so that a clear distinction especially between
the i=0-, i=1- and i=2-LP subbranches in real-space is hardly
possible, also because of the small energy splitting between them.

\begin{figure}[!t]
\centering
\includegraphics[width=\linewidth]{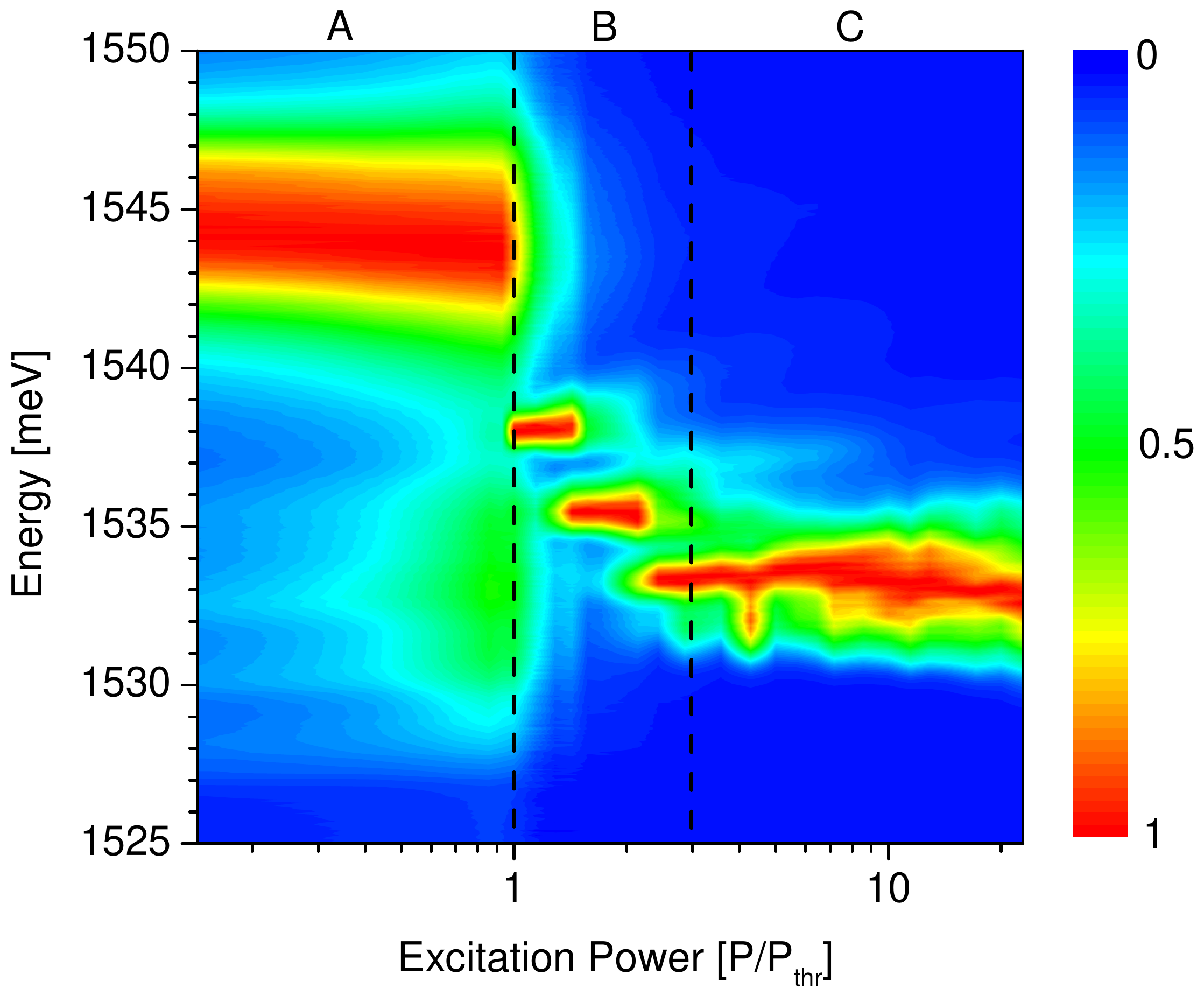}
\caption{(Color online) Power dependent spectra of the emission from
the photonic wire. Three different excitation regimes are indicated
by the labels A, B and C. 
Note: Emission was integrated over the full real-space image and normalized to one for each excitation power separately.}
\label{Fig4}
\end{figure}

\begin{figure}[!h]
\centering
\includegraphics[width=\linewidth]{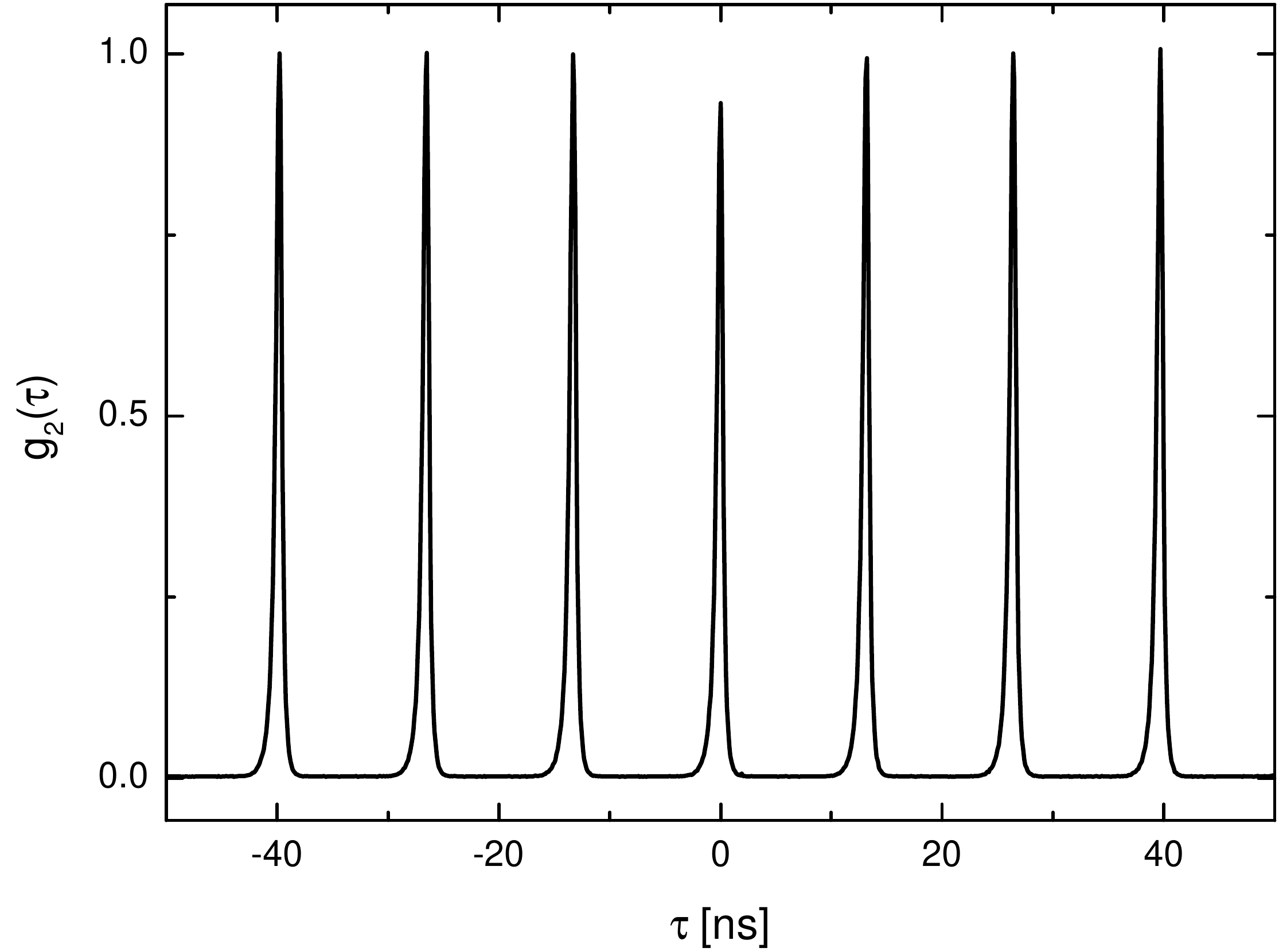}
\caption{Second order crosscorrelation between two polariton condensate modes at energies of $1532.6~\mbox{meV}$ and $1534.7~\mbox{meV}$. Excitation power amounts to $P=1.8P_{thr}$.}  
\label{Fig5}
\end{figure}

Fig. \ref{Fig4} summarizes the power dependent spectra for
increasing excitation power, divided into three different regimes.
Below threshold (regime A) the strongest emission comes from the
QWE, for intermediate excitation powers in regime B the main
emission is shifted from the i=4 LP subbranch to the LP i=0-, i=1-
and i=2-subbranches, whereas for high excitation power clearly above
threshold (regime C) the emission energy remains about constant.
In addition we have performed cross-correlation measurements using a HBT setup in the excitation regime  B between PoCos of different subbranches. For investigations in this regime near threshold the streak camera setup cannot be used due to the low duty cycle which is limited to 130 frames per second by the CCD camera.\cite{Assmann2010} The temporal resolution of our HBT setup is in the order of $500~\mbox{ps}$. Fig.~\ref{Fig5} shows a typical crosscorrelation measurement between two condensate modes at energies of $1532.6~\mbox{meV}$ and $1534.7~\mbox{meV}$, respectively. Clearly, antibunching at $\tau=0$ is observed. This indicates mode competition between PoCos in different subbranches. A similar antibunching effect was observed in Ref. \onlinecite{Ksudo2013} between two degenerate orbital states in a
honeycomb lattice potential which was attributed to stochastic
formation of different PoCos.

\subsection{First-order spatial coherence}

For spatially resolved investigation of the PoCo coherence properties
we choose high excitation power levels within regime C, where
pronounced propagation effects along the wire are observed.

\begin{figure}[!t]
\centering
\includegraphics[width=\linewidth]{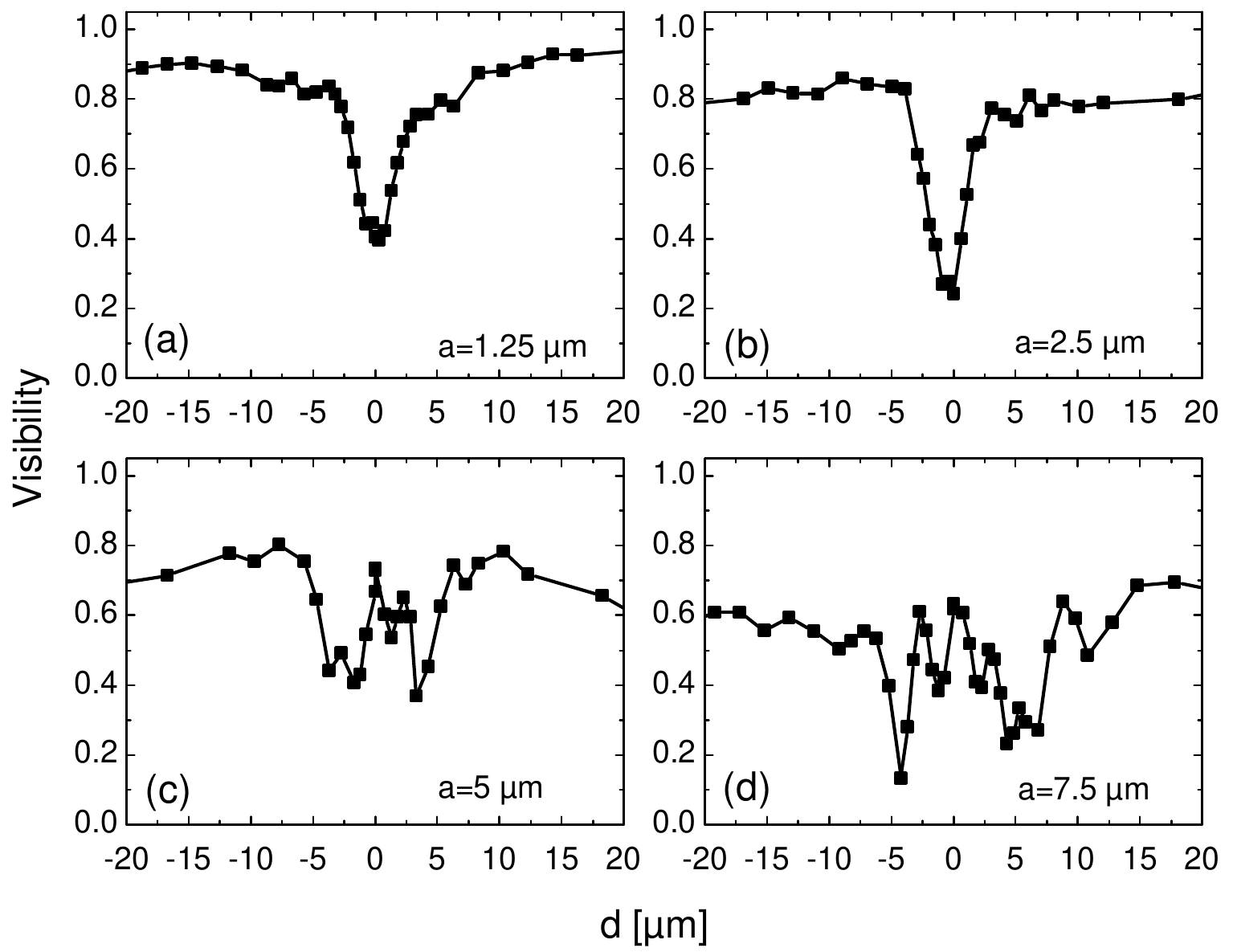}
\caption{Measured visibility $V(d)$ as function of the location $d$
of the double slit center for different slit separations $a$. The
energy of the emission is centered at $1532~\mbox{meV}$, the
excitation power is $P=14.3P_{thr}$. $d=0$ corresponds to the
situation of the double slit placed symmetrically with respect to
the excitation laser spot.} \label{Fig6}
\end{figure}

In Fig. \ref{Fig6} the dependence of the visibility on the probed
location of photonic wire is presented for an emission energy of
$1532~\mbox{meV}$ at an excitation power $P=14.3P_{thr}$. For
$|d|>10~\upmu\mbox{m}$ far away from the center of the excitation
laser spot, the visibility is more or less constant and shows the
expected monotonous behavior as observed elsewhere
\cite{Deng2007,Lai2007,Wertz2010}: The visibility
increases from roughly $0.6$ to $0.9$ when decreasing the slit
separation from $a=7.5~\upmu\mbox{m}$ to $a=1.25~\upmu\mbox{m}$.
However, in the vicinity of the laser spot around
$d=0$ a drastic decrease of the visibility becomes
evident for slit separations of $a=1.25~\upmu\mbox{m}$ and
$a=2.5~\upmu\mbox{m}$. The FWHM in both cases is approximately
$a=3.5~\upmu\mbox{m}$ which is in the order of the excitation laser
spot size. For the cases of $a=5~\upmu\mbox{m}$ and
$a=7.5~\upmu\mbox{m}$, no pronounced minimum of visibility at
$d=0$ is observed. Instead two pronounced minima
located symmetrically relative to $d=0$ are seen. In addition the
distance between the minima matches with the slit separation $a$. Therefore, the observation of the minima corresponds to the situation where the spatial coherence between PoCos located at the excitation laser spot and PoCos located $a=5~\upmu\mbox{m}$ and $a=7.5~\upmu\mbox{m}$, respectively, apart from the excitation spot is probed.

We tentatively assign the reduced spatial coherence around the laser
spot to interaction between condensed polaritons and the thermalized
reservoir of excitons localized around the excitation laser spot as
suggested in Refs. \onlinecite{Wertz2012,Wouters2010}. Recently, there was
a claim for observation of the detrimental effect of uncondensed
polaritons on the spatial coherence.\cite{Spano2012} In this study a
2D PoCo was created under the optical parametric oscillation
excitation scheme and the spatial coherence was compared between
phase-matching condition and excitation energy slightly shifted out
of phase-matching. In the latter case spatial coherence was found to
be decreased which was attributed to the detrimental effect of
uncondensed polaritons on the spatial coherence. However, the
decrease of coherence in this report might also be explained as
consequence of a lower density of the PoCo \cite{Manni2011a} in the
case of phase mismatch of the excitation laser. In our experiment we can rule out this explanation, as we observe similar polariton densities at the pump spot and $20~\upmu\mbox{m}$ apart in our real space spectra, whereas the visibility is $V(0~\upmu\mbox{m})=0.4$ at the pump spot and $V(20~\upmu\mbox{m})=0.9$ [Fig. \ref{Fig6} (a)].

\subsection{Spatially resolved measurement of second-order time correlation}

To substantiate our interpretation of the results of the double slit
experiment we additionally measure the $g_2(\tau)$-function
spatially resolved using the correlation streak-camera technique. To that end we place the excitation
laser at the edge of the photonic wire and image the emission of the
sample centered at $1530~\mbox{meV}$ onto the entrance slit of the
streak camera. Under this condition condensate
emission occurs at lower wavevectors compared to excitation in the
wire center. Thus, the intensity of the strongest emission feature
is redshifted compared to the power-dependent spectra shown in Fig.~\ref{Fig4}.

\begin{figure}[!t]
\centering
\includegraphics[width=\linewidth]{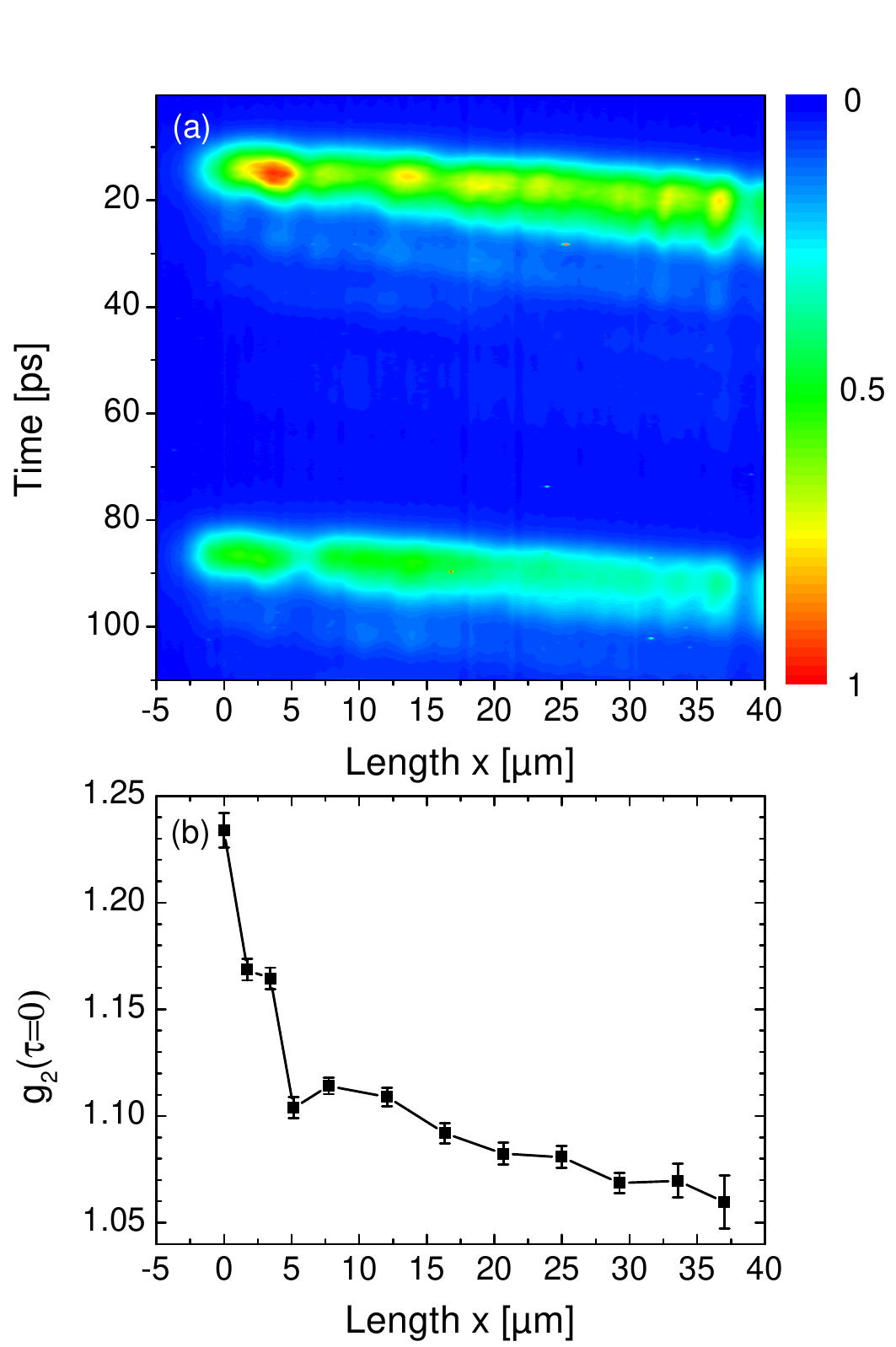}
\caption{(Color online) (a) Time-resolved spatial distribution of
the PoCo emission centered at $1530~\mbox{meV}$. Both profiles arise
from the same excitation pulse, but are delayed in time by
$72.5~\mbox{ps}$ due to different optical path lengths (see Sec. II for
explanation). The excitation power amounts to $P=30.4P_{thr}$. (b)
Measured $g_2(\tau=0)$ with respect to the position on the photonic
wire.} \label{Fig7}
\end{figure}

In Fig. \ref{Fig7} (a) the time-resolved spatial distribution of the
PoCo is shown. Here, $0~\upmu\mbox{m}$ indicates the location of the
excitation laser spot. Using a beamsplitter, we image the emission
from the photonic wire twice on the entrance slit of the streak
camera with a relative time delay of $72.5~\mbox{ps}$ to avoid photon
reconstruction errors for $\tau=0$ as outlined in Sec. II. From
this image one can deduce a group velocity of
$4.5~\upmu\mbox{m}~\mbox{ps}^{-1}$ in accordance with Ref. \onlinecite{Wertz2012}.
Individual second order correlations are measured collecting signal
from a region of the photonic wire with $3~\upmu\mbox{m}$ extension
using a vertical slit. Subsequently this
region is shifted along the wire. Fig. \ref{Fig7} (b) shows the
result of these spatially resolved measurements of ${g_2(\tau=0)}$,
the correlation function for simultaneous arrival of two photons.

We clearly see a bunching of photons emitted from the center of the
excitation laser spot, reflected by values increased above unity,
${g_2(\tau=0)=1.23}$. ${g_2(\tau=0)}$ decreases significantly within
$5~\upmu\mbox{m}$ down to ${g_2(\tau=0)=1.10}$. Further on, a slight
decrease down to ${g_2(\tau=0)=1.06}$ for a distance of
$37~\upmu\mbox{m}$ from the excitation laser spot is observed.
Whereas ${g_2(\tau=0)=1}$ reflects a Poissonian statistics and
therefore a coherent photon source, increased values of
$g_2(\tau=0)$ indicate a deviation from such a distribution and
hence a decreased coherence. 

\begin{figure}[!t]
\centering
\includegraphics[width=\linewidth]{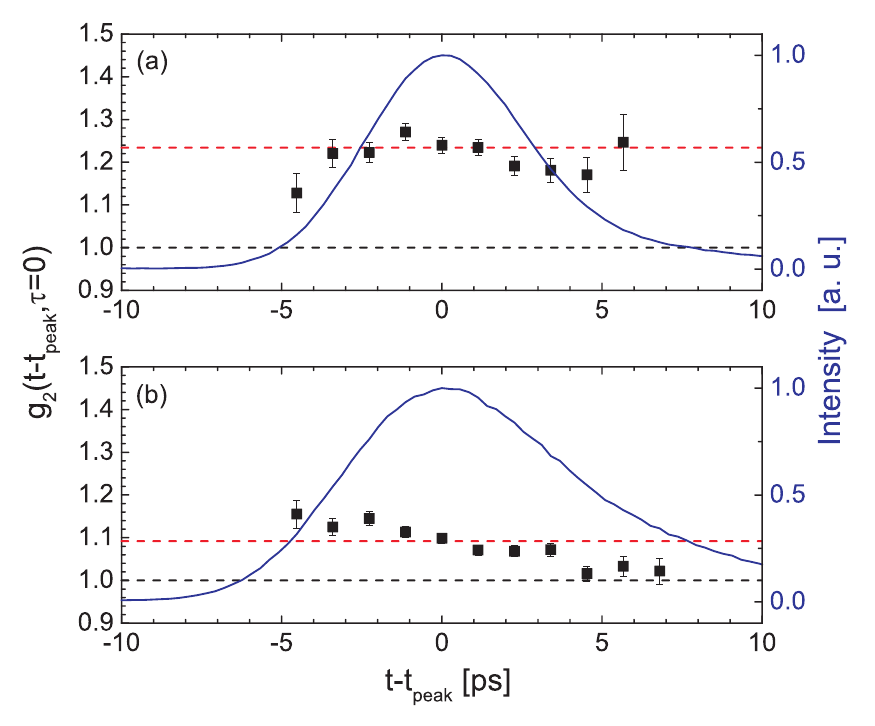}
\caption{(Color online) Time evolution of second-order photon-
correlation function $g_2(t,\tau=0)$ (black squares) compared to the normalized
output intensity (blue solid line) of the polariton condensate for $x=0~\upmu\mbox{m}$ (a) and $x=16.3~\upmu\mbox{m}$ (b). The red dashed line indicate the time averaged value $g_2(\tau=0)$ presented in Fig. \ref{Fig7}, the black dashed line the limiting case for coherent light. Note: Only times $t$ with a sufficient signal-to-noise ratio have been considered for the evaluation of $g_2(t,\tau=0)$.}
  \label{Fig8}
\end{figure}

We have additionally analyzed the $g_2(t,\tau=0)$-function for different positions along the photonic wire (Fig.~\ref{Fig8}). At the location of the excitation laser spot we can see only small fluctuation around the mean value of ${g_2(\tau=0)=1.23}$ within the emission pulse of the polariton condensate [Fig. \ref{Fig8}~(a)]. Interestingly, far away from the excitation laser the situation is different [Fig. \ref{Fig8}~(b)]: We observe a monotonous decrease of $g_2(\tau=0)$ towards 1 within the pulse which demonstrates the recovery of a coherent light emission when no reservoir of background carriers is present. 
Therefore this experiment additionally corroborates our interpretation of a decreased coherence of the PoCo when background carriers are present. 

We have also considered the possibility that the high $g_2$-values observed at the pump spot can be interpreted in terms of simultaneously detected thermal photons. However, an analysis of the input-output curve at the location of the excitation laser spot revealed an ratio of roughly $3~\%$ of thermal photons and $97~\%$ of photons from the polariton condensate within the timewindow of roughly $10~\mbox{ps}$ of the polariton condensate emission [Fig. \ref{Fig8} (a)]. Even for the very unlikely case that every detected pair consisting of a thermal/coherent photon would contribute with $g_2(\tau=0)=2$, the overall value for $g_2(\tau=0)$ would be $1.06$ which is significantly lower as the average value $g_2(\tau=0)=1.23$ within the emission pulse from the polariton condensate at the excitation laser spot [Fig. \ref{Fig8} (a)]. Therefore we can exclude that our results can be explained in terms of simultaneously measured thermal photons.

A similar decrease of second-order coherence induced by interaction with a reservoir was recently observed for a photon BEC.\cite{Schmitt2014} One of the key findings of this report is the observation of an increased particle number fluctuation for decreasing condensate fraction with respect to the reservoir (excited dye-molecules in this study) which is evidenced by an increase of ${g_2(\tau=0)}$ up to values of $1.7$ for low condensate fractions. The high $g_2$-values observed are attributed to the grand-canonical ensemble conditions of the experiment when the condensate fraction is low and the particle exchange between the reservoir and the condensate is very effective.  
To our opinion, a similar effect is seen in our experiment: As the reservoir is mainly located within the excitation laser spot we have a gradient of low condensate fractions within the laser spot to high condensate fractions several $10~\upmu\mbox{m}$ away from the excitation laser.

\section{conclusion and outlook}
In conclusion, we have demonstrated the detrimental effects of
background carriers on the coherence properties of PoCos using Young's
double slit experiment and second order correlation measurement due to interaction between the reservoir and the PoCo.
We have also presented a technique to determine second order correlation spatially resolved, which should also allow one to measure second order cross-correlations of PoCos in space. This could pave the way for the identification of event horizons exhibiting Hawking radiation.\cite{Solnyshkov2011,Gerace2012}

\section{acknowledgments}
The Dortmund group acknowledges support from the Deutsche Forschungsgemeinschaft (grants 1549/18-1 and 1549/19-1). We are grateful to Julian Fischer for initial sample characterization and measurements. Technical support in the fabrication of microwires by M. Emmerling and A. Wolf is acknowledged.

\end{document}